\documentclass[a4paper]{aa}

\usepackage{float,psfig}
\def \src {XMMU\thinspace{J183225.4-10364}}

\def \nh {N${\rm _H}$}

\def \hcm {\hbox {\ifmmode $ atom cm$^{-2}\else atom cm$^{-2}$\fi}}
\def \arcmin {\hbox{$^\prime$}}
\def \arcsec {\hbox{$^{\prime\prime}$}}

\def\approxgt{\mathrel{\hbox{\rlap{\lower.55ex \hbox {$\sim$}}
        \kern-.3em \raise.4ex \hbox{$>$}}}}
\def\approxlt{\mathrel{\hbox{\rlap{\lower.55ex \hbox {$\sim$}}
        \kern-.3em \raise.4ex \hbox{$<$}}}}

\begin{document}

\title{Discovery of an absorbed cluster of galaxies 
(XMMU\thinspace{J183225.4-103645}) close to the Galactic plane with 
XMM-Newton}

\author{J. Nevalainen\inst{1} 
\and   D. Lumb\inst{1}
\and   S. dos Santos\inst{2}
\and   H. Siddiqui\inst{1}
\and   G. Stewart\inst{2}
\and A. N. Parmar\inst{1}}

\offprints{J. Nevalainen (jnevalai@astro.estec.esa.nl)}

\institute{
           Astrophysics Division, Space Science Department of ESA, ESTEC,
           Postbus 299, NL-2200 AG Noordwijk, The Netherlands
\and       Department of Physics and Astronomy, University of Leicester, 
           University Road, Leicester LE1 7RH, UK}

\date{Received ; Accepted: }

\abstract{During an XMM-Newton observation of the galactic supernova remnant 
G21.5-09 a bright, previously uncatalogued, source (XMMU~J183225.4-103645)
was detected 18\arcmin\ from G21.5-09. The European Photon Imaging Camera 
data inside 1\arcmin\ (180~h$_{50}^{-1}$ kpc) radius are consistent with 
a source at a redshift of $0.1242 \pm ^{0.0003}_{0.0022}$ with an optically 
thin thermal spectrum  of temperature  
$5.8 \pm 0.6$~keV and a metal abundance of $0.60 \pm 0.10$ solar.
This model gives a 2 -- 10 keV luminosity of 
${\rm 3.5^{+0.8}_{-0.4}  \ h_{50}^{-2} \ 10^{44} \ erg~s^{-1}}$.
These characteristics, as well as the source extent of 
2\farcm0 (350 ${\rm h_{50}^{-1}}$~kpc), 
and the surface brightness profile are consistent with  
emission from the central region of a moderately rich cluster containing a 
cooling  flow with mass flow rate of ${\sim \! 400- 600} \ {\rm M_{\odot}}$~yr$^{-1}$.  The absorption is 
$(7.9 \pm 0.5) \ 10^{22}$~atom~cm$^{-2}$, 
5 times that inferred from low-resolution HI data but 
consistent with higher spatial resolution infrared 
dust extinction estimates.
XMMU~J183225.4-103645 is not visible in earlier ROSAT observations due to 
high amount of absorption. This discovery demonstrates the capability of 
XMM-Newton to map the cluster distribution close to the Galactic plane, where
few such systems are known. The ability of XMM-Newton to determine cluster 
redshifts to 1\% precision at z = 0.1 is especially important in optically crowded and 
absorbed fields such as close to the Galactic plane, where the
optical redshift measurements of galaxies are difficult. 
}

\maketitle

\markboth{Galactic plane galaxy cluster discovered with XMM-Newton}
         {Galactic plane galaxy cluster discovered with XMM-Newton}

\section{Introduction}

Studies of large scale structure in the Universe have been extended with the 
advent of X-ray surveys.  Clusters trace the deepest gravitational potential 
wells, and their X-ray emission, which traces out the hot phase of 
inter-galactic gas, is less dependent on foreground contamination effects 
than optical cluster identification techniques. In addition, optical searches 
for clusters of galaxies have historically been forced to avoid a wide band of sky
centered on the Galactic plane. The obscuration of $\sim$25\% of the whole sky 
by the Galactic plane is problematic for the understanding of the dynamics in
the local Universe, where the whole-sky map of the large scale structure is 
essential (Kraan-Korteweg \& Lahav 2000).
This is particularly unfortunate from the point of view of 
mapping out structure associated with the Great Attractor region
(l ${\sim \! 320^{\circ}}$, b ${\sim \! 0^{\circ}}$)
More recently a number of searches employing techniques in different 
wavebands have attempted solve this problem (see e.g., Schneider et al. 1997).

With X-ray searches, the limiting factor is no longer extinction due to dust 
and confusion from stellar sources, but the amount of photo-electric 
absorption, ${\rm N_H}$. The ${\rm N_H}$ and visual extinction rise roughly 
proportionally towards the Galactic plane, but because XMM-Newton has a lot of 
effective area above energies 2 keV, where the absorption is negligible, the 
photon loss is much smaller with XMM-Newton, compared to optical band. Therefore, 
XMM-Newton survey for clusters should be able to reach closer to the plane than 
traditional survey techniques.
Ebeling et al. (2000) describe such a survey using data from the ROSAT All-Sky
Survey (RASS). Initially, they examined sources in the ROSAT Bright Source Catalogue,
where targets were selected with fluxes 
$>$$3  \ 10^{-12}$~erg~cm$^{-2}$~s$^{-1}$,
$|b|$ $< 20^{\circ}$,
and hard spectral colors. From follow-up studies they obtained 73
spectroscopically confirmed clusters, of which 58 were previously unknown.  
Nevertheless, the ROSAT pass-band of 0.1--2.5~keV limited the 
sensitivity to energies  where absorption is high, even for modest
galactic extinctions $A_v \sim 1$. The newly commissioned
{\it Chandra} and XMM-Newton X-ray observatories have high sensitivities 
extending to 10 and 12 keV, respectively. 
Therefore, {\it Chandra} and XMM-Newton have additional potential to discover 
highly-absorbed objects, with equivalent optical extinctions  $A_v > 20$
 compared to ROSAT.
In this work we demonstrate the capability of 
detecting  clusters behind the central regions of the Milky Way.

We use $H_{0} \equiv 50 \ {\rm h_{50}}$ km s$^{-1}$ Mpc$^{-1}$, 
$\Omega_{m} = 1$, $\Omega_{\Lambda} = 0$, 
and consider uncertainties and significances at 90\% level throughout the 
paper, unless stated otherwise.

\section{Observation}

The XMM-Newton payload comprises 3 identical X-ray reflecting telescopes, the
EPIC (European Photon Imaging Camera) imaging spectrometers, together with 
two reflection grating spectrometers and a co-aligned optical/UV monitor. The
EPIC complement includes 2 conventional MOS CCD detectors (Turner et al. 2001)
and a PN CCD instrument (Str\"uder et al. 2001). The former are placed behind 
the mirror systems containing grating arrays, the latter in the unobscured 
telescope. For more details of XMM-Newton see Jansen et al. (2001).
In total the 3 EPICs provide $>$2500~cm$^{2}$ of collecting area at 1.5~keV. 
The mirror systems offer an on-axis full-width half maximum (FWHM) angular 
resolution of 4--5\arcsec\ and a field of view (FOV) of 30\arcmin\ diameter. 

The location of the reflection grating array behind two of the mirror systems
breaks the rotational symmetry of the collecting area vignetting. To confirm 
the ground calibration of this effect, observations were made of a number of 
sources at on- and off-axis positions. The compact galactic supernova remnant
G21.5-09 was chosen for these measurements because it is time invariant,
has a compact core with a spatial scale not much larger than the XMM-Newton
point spread function (PSF), and a simple absorbed power-law spectrum (see 
Warwick et al. 
2001 for the analysis of XMM-Newton observations of G21.5-09). During one of 
these off-axis observations on 2000 April 11 the source discussed here was 
discovered serendipitously.

The new source is located on the opposite side of the FOV from G21.5-09 at an
off-axis angle of 8\arcmin\, at 
RA = 18${\rm ^{h}}\,$32${\rm ^{m}}$\,25.4${\rm ^{s}}$,  
dec. = $-10{\rm ^{\circ}}$\,36\arcmin\,45\arcsec\ 
(corresponding to galactic coordinates 
l = 21.3${\rm ^{\circ}}$, b = $-0.7{\rm ^{\circ}}$) 
with a 5\arcsec\ uncertainty in the absolute coordinates. At this off-axis 
angle the PSF of the XMM-Newton optics is barely degraded from the on-axis 
value. The source is designated \src. Exposures of 29~ks were taken with the 
0.2--10~keV EPIC in full frame mode using medium thickness filters. 
The event lists produced by the XMM-Newton Science Analysis Software (SAS 5.0)
tasks {\sc emchain} and {\sc epchain} were subsequently filtered using a third 
SAS task, {\sc xmmselect}. For the MOS only X-ray events corresponding to 
patterns 0--12 were selected, while for the PN, only pattern 0 (single pixel)
events were selected. Known hot, or flickering, pixels and electronic noise 
were rejected using the SAS.

We examined the effect of particle flares by extracting a light curve of
all the CCDs in the $>$10 keV band. The light curve shows several short peaks, 
and a large flare towards the end of the observation. We rejected time
bins using several different count rate criteria, ranging from no rejection at
all to a very low level (0.4 count s$^{-1}$ for the PN and 0.2 count s$^{-1}$
for the MOS), which reduced the exposure time by 50\%. Fitting the 
corresponding PN + MOS spectra revealed, that the fit parameters are 
virtually independent of the rejection level. However, to be safe, we chose 
to reject data obtained during the large flare close to the end, which 
reduced the exposure time to 26 ks.

\section{Spatial Analysis}
\label{spatext}

\begin{figure}
\centerline{\psfig{figure=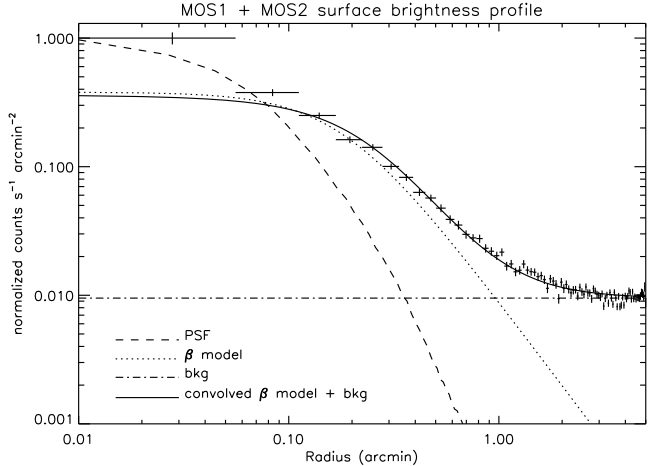,width=9.0cm,angle=0}}
\caption[]{The vignetting corrected MOS1 + MOS2 surface brightness profile 
centered at the emission peak in the 1--8 keV band with $1\sigma$ 
uncertainties (crosses). The PSF (dashed line) is much narrower than the 
data, indicating that the source is extended. A best-fit $\beta$ model to 
the 0\farcm06 -- 5\farcm0 profile data (dotted line) convolved with the PSF + background
(dot-dash), is shown as a solid line}
\label{surfprof_fig}
\end{figure}

To study the spatial extent of the source, we used the SAS tool {\sc eexpmap}
to create exposure maps in the 1--10~keV energy range for the PN observation,
and in 1--8~keV range for MOS1 and MOS2.
The exposure maps include effects of the spatial dependence 
of the quantum efficiency, filter transmission and mirror vignetting. 
We used the images to derive radial surface brightness profiles
for PN, and a combined one for MOS1 + MOS2 (Fig.~\ref{surfprof_fig}).

For comparison, we used {\sc SciSim} V2.1 to produce PN and MOS images of a 
point source at the same location and with a spectrum similar to that of 
{\src} (i.e. a {\sc mekal} spectrum with T = 6 keV and absorbed by 
${\rm N_{H} = 8 \ 10^{22}}$~atom~cm$^{-2}$, see below). The {\sc SciSim}
mirror model contains relevant geometric and metrology data that was verified
by the ground calibration, and subsequently by in-flight calibration 
measurements (Aschenbach et al. 2000). We extracted a radial surface 
brightness profile of the simulated image and thereby obtained the PSF 
corresponding to our observation. Comparison of the surface brightness 
profile data with PSF shows that the source is clearly extended
(Fig.~\ref{surfprof_fig}). With MOS data, we can detect the source out to 
a 2\farcm0\ radius with $3\sigma$ confidence. Using the best-fit reshift 
(see below) this radius corresponds to 350 h$_{50}^{-1}$~kpc. In the following,
we use this conversion between angular and physical radii.
Beyond a 3\arcmin\ radius the surface brightness reaches a nearly constant 
value, indicating that the sky and particle background dominates over the 
source emission at these radii. 

We determined the background level from the data at large radii (beyond 3 
\arcmin\ ) and fitted 
the MOS data at radii of 0\farcm06 -- 5\farcm0 and the PN data at radii
0\farcm1 -- 4\farcm0 with a $\beta$ model (Cavaliere \& Fusco-Femiano 1976) 
and background, convolved with the PSF. A single $\beta$ model provides
a reasonable description of both data sets with $\beta = 0.53 \pm 0.02 $ and
$r_{{\rm core}} = 12\arcsec \pm 1\farcs2 =  35 \pm 4$ h$_{50}^{-1}$~kpc, except in the 
central $\sim$ 4\arcsec\ bin, where the count rate is significantly 
(by a factor of 3.0) above the model. The low value of the core radius
is more suggestive of values obtained for rich cluster cooling flows than 
the larger scale ambient cluster emission. We have therefore also fit a 
two-component model with the second component having fixed values of 
$\beta$ = 2/3 and $r_{{\rm core}}$ = 250 h$_{50}^{-1}$~kpc, 
representative of other typical clusters. The best fit cooling flow 
component has values 
$\beta = 0.74 \pm 0.07$ and $r_{{\rm core}} = 40 \pm 10$ h$_{50}^{-1}$~kpc, typical
of a large sample of clusters (Mohr et al. 1999).
We find that $\sim 70\%$ of the counts within 1\arcmin\ 
can be attributed to a `cooling flow' component with the residual coming 
from the more extended component.

\begin{figure}
\begin{center}
\vbox{
\psfig{figure=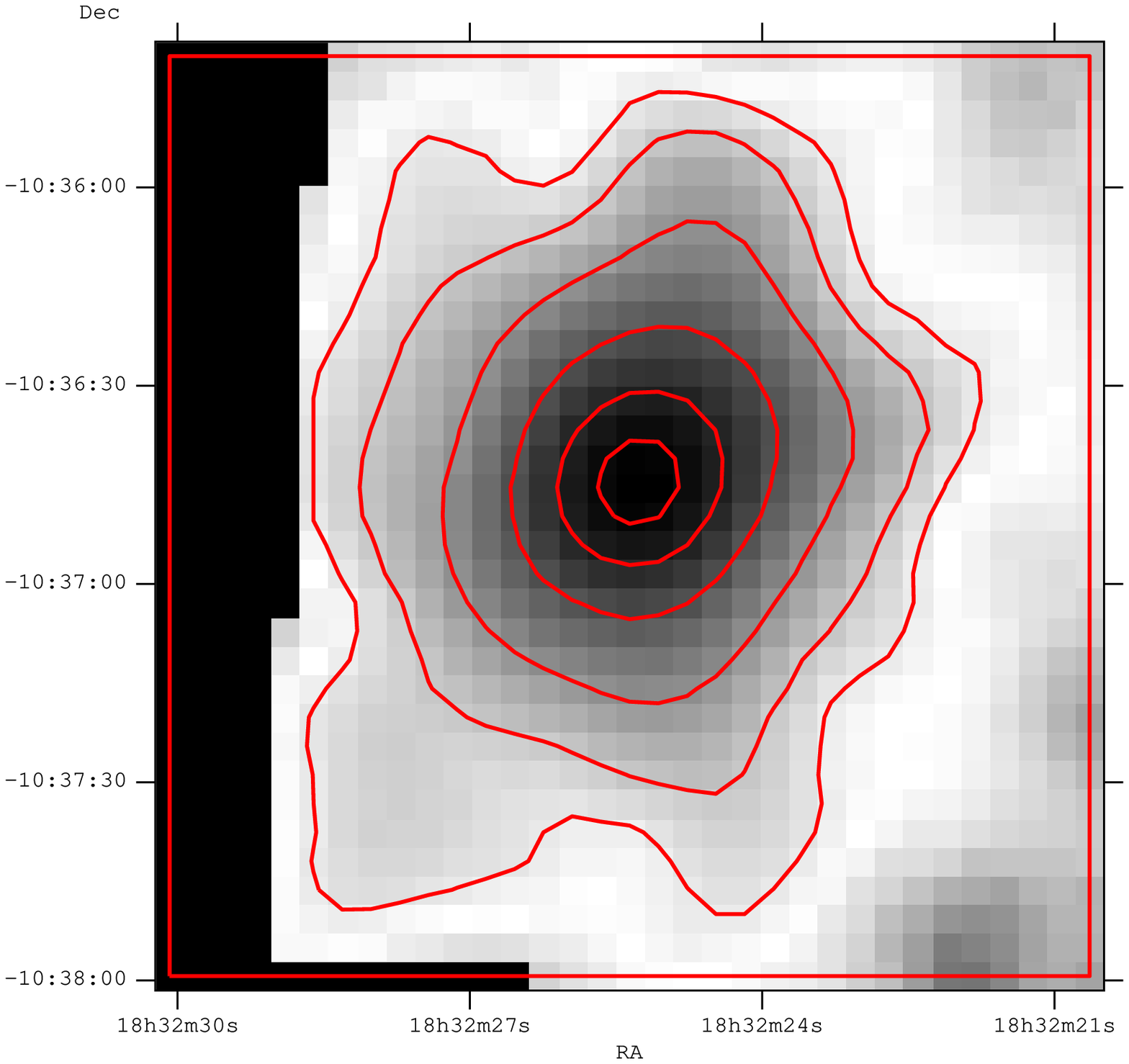,width=8.5cm,angle=0.}
\psfig{figure=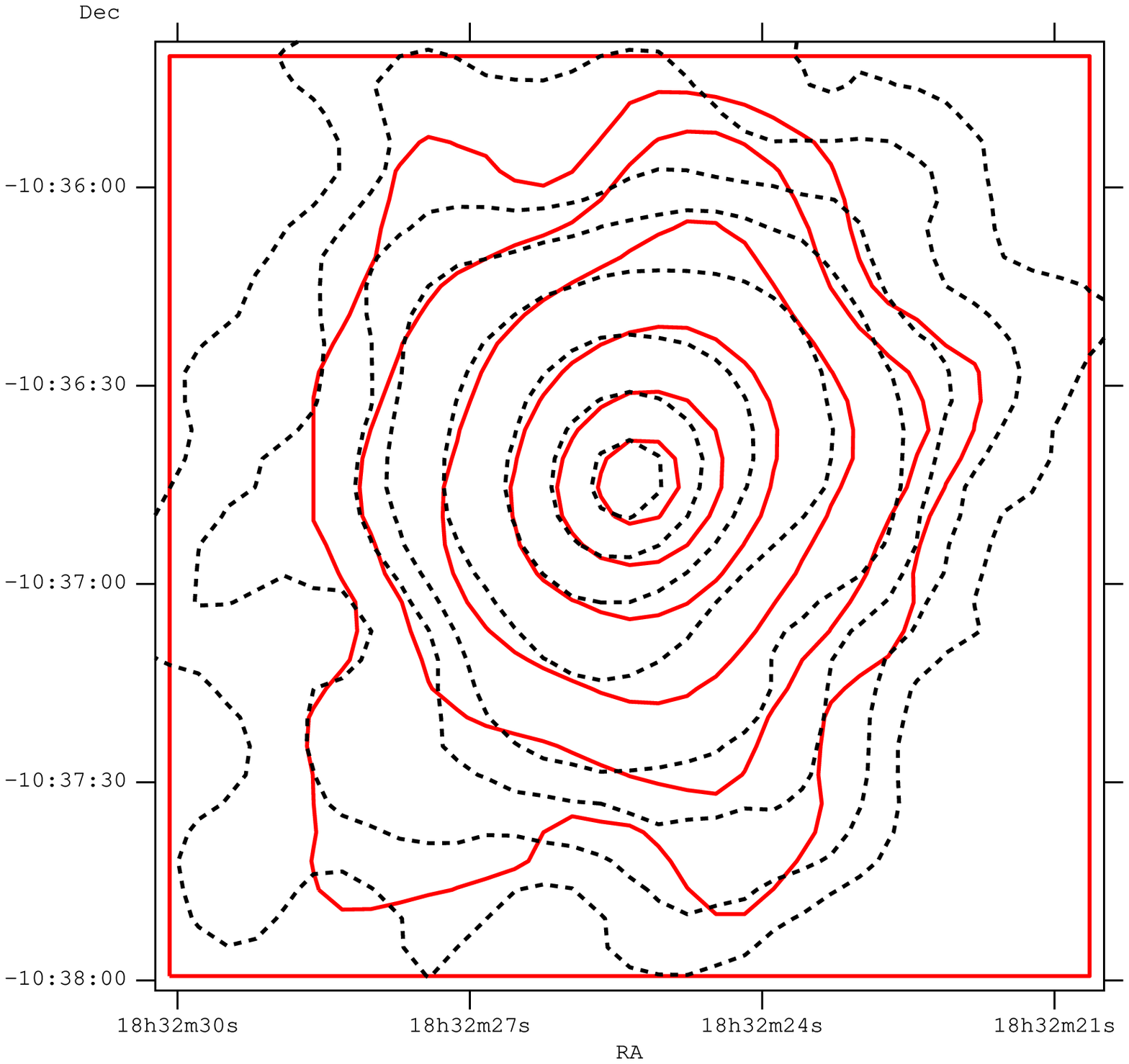,width=8.5cm,angle=0.}
}
\end{center}
\caption[]{The upper panel shows the vignetting corrected 1--10 keV PN 
image of {\src}. The pixel size is 4\farcs35 , about the size of the 
FWHM of the PSF. The image has been smoothed using a 
Gaussian with $\sigma$ = 9\arcsec. The contours show the brightness 
levels of 91\%, 69\%, 43\%, 24\%, 15\% and 12\% of the peak value.
The lower panel shows the same contours as in upper panel, together with
brightness contours from 1 -- 8 keV MOS2 image, smoothed by 2 pixels 
(= 7\arcsec ). The levels are the same as for PN, except 
that for MOS2 additional 9\% and 7\% contours are shown}
\label{pnimage_fig}
\end{figure}

The exposure corrected PN image (Fig.~\ref{pnimage_fig}) does not exhibit
strong distortions from azimuthal symmetry and thus does not suggest 
a strong ongoing merger in the cluster. However, the data indicates 
small radial elongation structures between 0\farcm5\ -- 1\arcmin\ and 
central 0\farcm5\ elongation with an axial ratio of $\sim$ 0.85.
The spokes in the mirror produce a 16-fold symmetric pattern in surface 
brightness at the spatial scales smaller than the pixel size used and thus
cannot produce the observed features. 
Due to the small number of counts in the pixels at 1\arcmin\ distance from 
the center, we cannot derive strong 
conclusions on the spatial structure. However,
the MOS2 brightness contours seem to confirm the existence of the deviations
from the spherical symmetry, decreasing the probablility of them being noise.
In cooling flow scenario, the elongated radial structures could result from
enhanged density due to patchy infalling gas. Similar behavior
is visible in Chandra data of the core of cooling flow cluster 3C295 
(Allen et al. 2001). The central elongation could be due to a recent
merger, which is about to disrupt the cooling flow.

In order to obtain information on the spatial distribution of the spectral 
properties of the cluster gas, we produced a color map, where color is 
defined as vignetting corrected PN counts in the 3--10 keV energy range,
divided by the counts in the 1--3 keV energy ranges (Fig. 3).
The pixel size was chosen to be 26\arcsec\ so that each pixel 
contains at least 20 counts in both bands. The resulting standard 
Gaussian errors of the color derived from the raw counts are below 1 
in most pixels. Within 1\arcmin\ the color varies significantly, from 3 to 8.
The emission in the South-West quadrant is soft (color $\sim \! 3$), 
while the North-East quadrant is harder (color $\sim \! 3$--5). Between them,
aligned with the elongation axis of the total band brightness the emission is very 
hard (color $\sim \! 7$--8). If this is due to significant variation in \nh , 
the excess 
absorption would affect the soft band more and increase the color value.
This would lead to an anticorrelation between the soft band counts and the 
color, but this is excluded by the data.
This indicates that the variations are due to temperature variations in 
the cluster. It is interesting to note that Chandra imaging of a core of 
cooling flow cluster Perseus (Fabian et al. 2000) shows relatively well
defined regions of different temperature gas and strong temperature 
variations on very small angular scales, qualitatively similar to our findings
for {\src}. Such a structure may be produced by shock waves resulting from 
the collision of different infalling gas patches. Alternately, the variations 
in the color across the elongation axis of the cluster could also suggest a 
merger along the same axis, so that areas of significant hardening would be due
to heating due to ram-pressure.
  
\begin{figure}
\begin{center}
\vbox{
\psfig{figure=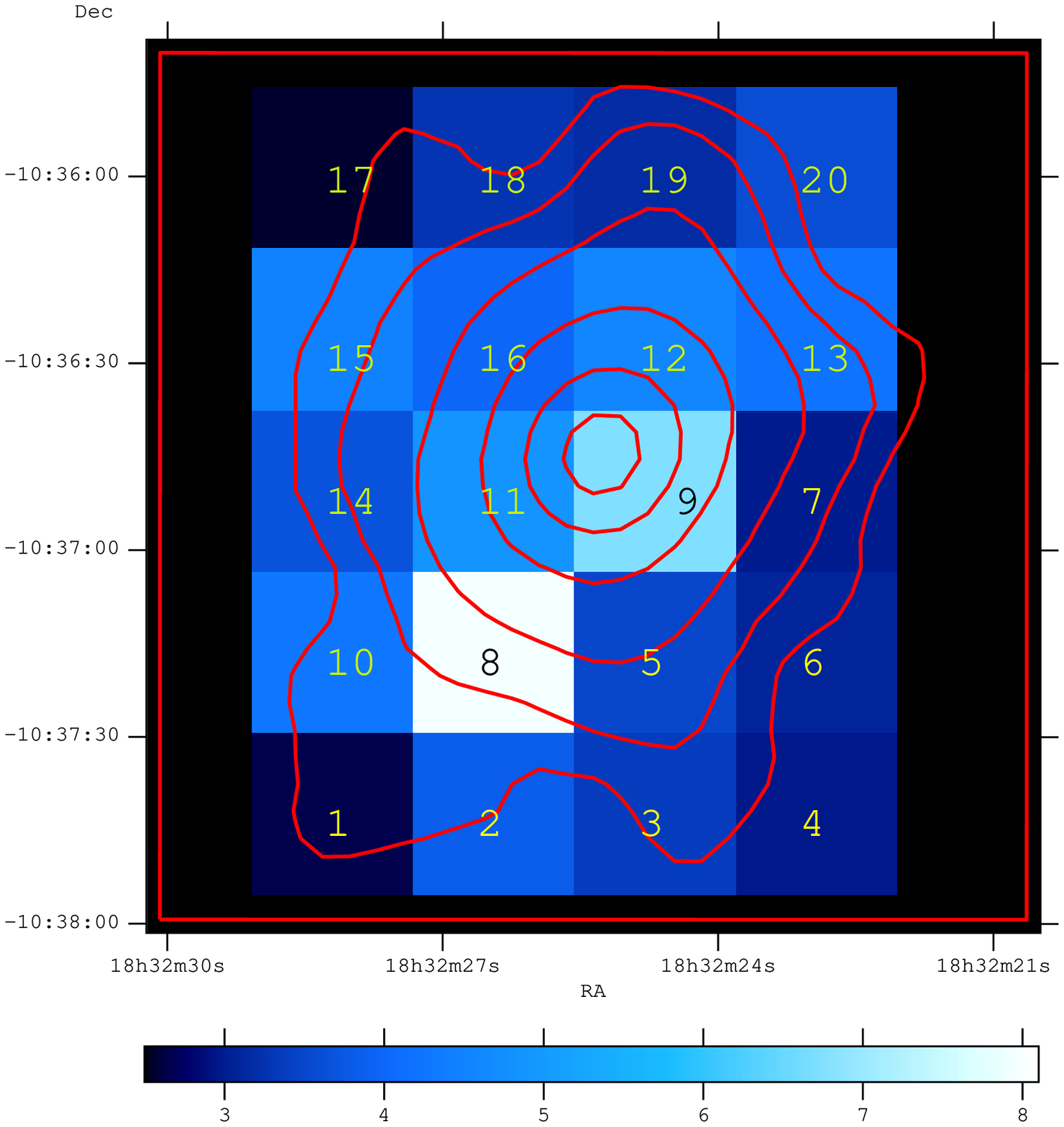,width=9.cm,angle=0.}
\psfig{figure=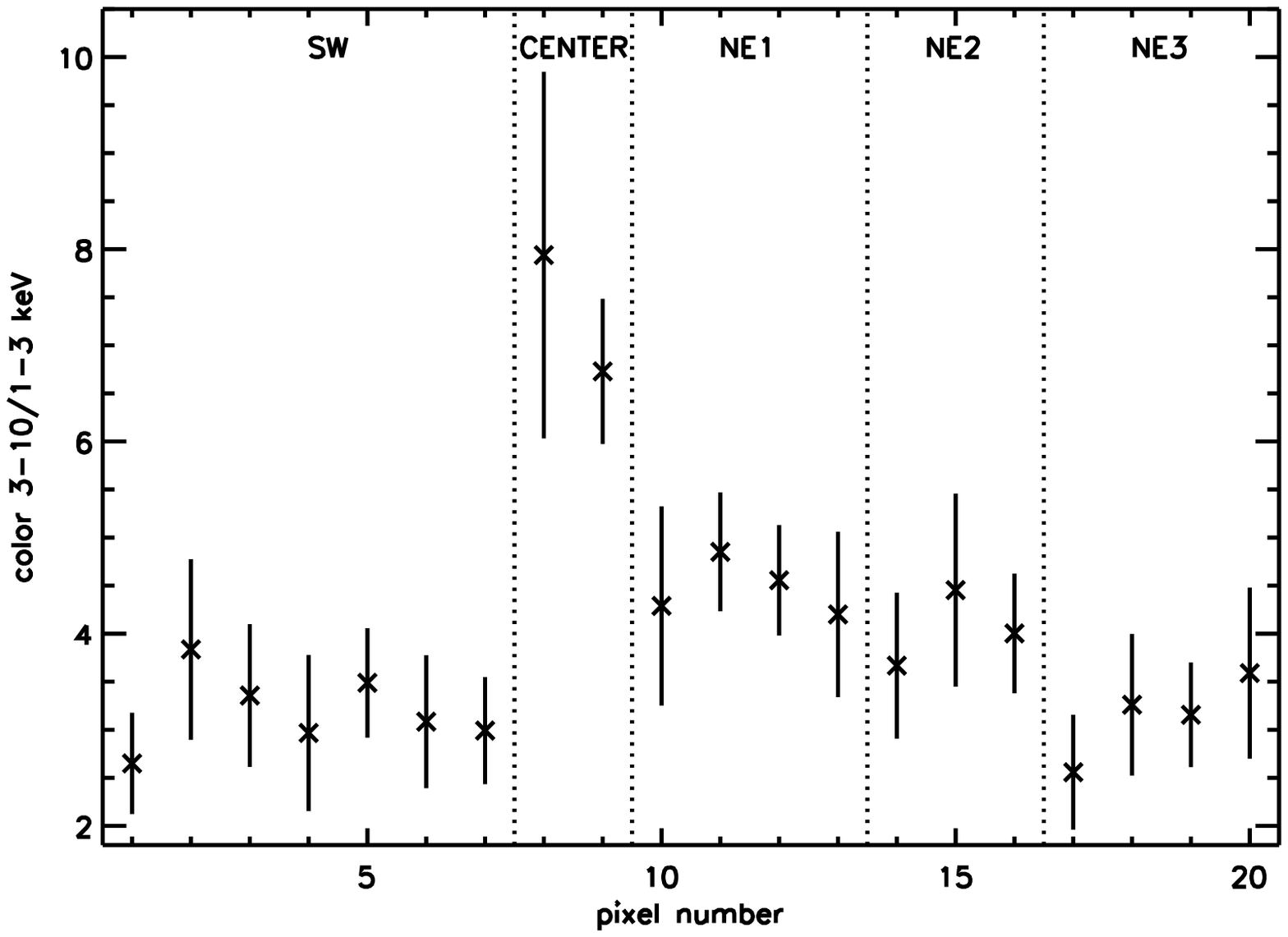,width=9.cm,angle=0.}
}
\end{center}
\caption[]{The upper panel shows the 3--10/1--3 keV color map of {\src} 
obtained from PN data. The pixel size is 26\arcsec\ . The X-ray emission 
contours from 1--10 keV PN image are as in Fig. 2. The lower panel shows the color values with standard gaussian errors for each pixel. The dotted 
vertical lines separate the different regions going from the lower right 
corner of the image towards the upper left corner. The central hardening of 
the spectrum is clearly evident}
\label{colormap_fig}
\end{figure}

\section{Spectral Analysis}

Source counts were extracted from circular regions of 1\farcm1\ and
1\farcm2\ radius centered on \src\ for the PN and MOS-2 instruments.
In the MOS-1 instrument the source fell close to the gap between two of 
the outer CCDs of the
detector and thus an extraction radius of 0\farcm9\ was used.
The background extraction required extra care, 
because the particle induced fluorescent excitation of materials in XMM-Newton
produces spatially variable K line
contamination for the PN, and Al K and Si K line contamination for the MOS.
The intensity of these features varies with position within the FOV.
To study this, we extracted background spectra from the PN image
varying the extraction locations as a function of a distance from the source,
up to 10\arcmin.\ We also used locations at different off-axis angles
(up to 10\arcmin)\ to see if the radially-changing vignetting function 
produces a significant effect.
We fitted the data using these background spectra, and noticed that even 
though there were some differences in the PN spectrum at 8 keV (due to the Cu
line) and below 2 keV (due to the Al line) 
these made an insignificant effect to the fit parameters, compared to the 
statistical uncertainties. Thus, the choise of background extraction region 
is not critical in our case, but again, to be safe, we chose to use 
background regions close to the source and at similar off-axis angle in the 
subsequent analysis.

As suggested by Saxton \& Siddiqui (2000), we used the ready-made on-axis 
energy redistribution matrices epn\_fs20\_sY9.rmf, M1\_all.rmf and 
M2\_all.rmf for the PN, MOS-1 and MOS-2, respectively, assuming that the 
response function does not vary significantly across the CCDs. We generated 
the vignetting corrected auxiliary response files 
with the SAS tool {\sc arfgen 1.41.12}. In our fit, we excluded the strongly 
absorbed
low energy band (below 1 keV), where most of the signal consists of 
redistributed photons from higher energies, and used the energy ranges 
1--10~keV  for the PN and  1--8~keV for the MOS. The net count rates in the 
above bands are 
$0.094 \pm 0.002$, $0.049 \pm 0.002$, and $0.060 \pm 0.002$ s$^{-1}$ for
the PN, MOS-1 and MOS-2, respectively.
We rebinned the data so that each bin included at least 50 counts.
The photoelectric absorption
cross sections of Morrison \& McCammon (\cite{m:83}) and the
solar abundances of Anders \& Grevesse (\cite{a:89}) were used throughout

\begin{figure}
\centerline{\psfig{figure=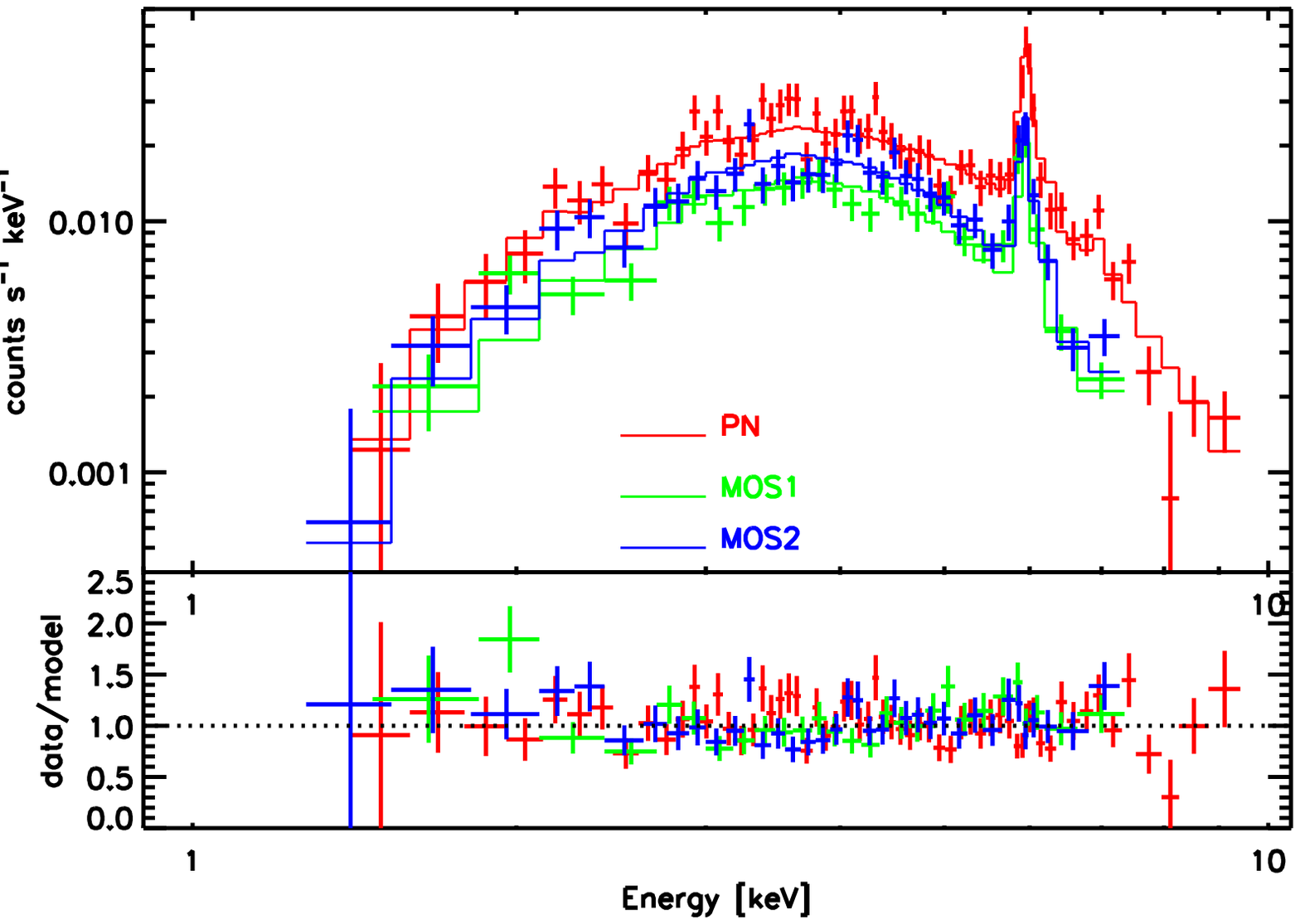,width=9cm,angle=0}}
\caption[]{The best-fit absorbed {\sc mekal} model convolved with the 
instrumental responses 
of the PN, MOS-1 and MOS-2 are shown as solid lines, together with the data and
1$\sigma$ uncertainties (crosses).
A redshifted Fe emission feature at 6.0 keV is clearly evident. 
The lower panel shows the ratio of the data and the best-fit model}
\label{spectrfit_fig}
\end{figure}

We fitted the spectrum using the absorbed {\sc mekal} 
collisionally ionized thermal plasma model in XSPEC v11.0.1. The 
${\rm N_{H}}$ and the redshift were allowed to vary as 
well as the relative normalizations between the PN, MOS-1 and MOS-2 
instruments.
We obtained the best fit with parameters
${\rm N_{H} = (7.9 \pm 0.5) \ 10^{22}}$~atom~cm$^{-2}$, 
T = $5.8 \pm 0.6$~keV,
metal abundance of $0.60 \pm 0.10$ in Solar units and a redshift of 
$0.1242^{+0.0003}_{-0.0022}$ (see Fig.4 and Table~1).
The fit is formally acceptable at a 90\% confidence level with 
a $\chi^2$ of 140.1 for 120 degrees of freedom (dof).
The residuals show that in PN data at 8 keV the data bins are significantly 
below the model, indicating the effect of the variable Cu K $\alpha$ line.

\begin{table}
\caption[]{The best fit values for the serendipitious cluster using 
{\sc wabs * mekal} model in XSPEC. The uncertainties are given at 90\% 
confidence level. The spectra are extracted 
within  1\farcm1, 0\farcm9 and 1\farcm2 radii for the PN, MOS1 and MOS2, 
respectively. The energy ranges used for the fits are 1--10 keV for PN and 
1--8 keV for 
MOS-1 and MOS-2}
\begin{flushleft}
\begin{tabular}{ll}
\hline\noalign{\smallskip}
                                                          & PN + MOS1 + MOS2    \\
\hline\noalign{\smallskip}
T (keV)                                                   & $5.8 \pm 0.6$    \\
Metallicity (solar)                                       & $0.60 \pm 0.10$  \\
\nh  $(10^{22} \ {\rm atom} \ {\rm cm}^{-2}$)             & $7.9 \pm 0.5$    \\
redshift                                                  & $0.1242 \, \pm \,  ^{0.0003} _{0.0022}$ \\
$\chi^2$/dof                                              & 140.1/120         \\
${\rm F_{2-10}  \ (10^{-12} \ erg~s^{-1}~cm^{-2})}$       & $4.9 \, \pm \,  ^{1.0} _{0.6}$ \\
${\rm L_{2-10}   \ (h_{50}^{-2} \ 10^{44} \ erg~s^{-1})}$ & $3.5 \, \pm \,  ^{0.8} _{0.4}$  \\
\noalign{\smallskip\hrule\smallskip}
\end{tabular}
\end{flushleft}
\label{spectrfit_tab}
\end{table}

Our redshift measurement depends critically on the centroid energy of the 
Fe K$_{\alpha}$ line, and thus on the accuracy of the instrument gain. As 
shown in Str\"uder et al. (2001), the long term monitoring of the centroid of
the Mn K$_{\alpha}$ line of the internal PN radioactive spectral calibration 
source shows that it
has been stable to within 5 eV between 2000 January and September. 
This shows that in the long term the gain is well defined, at least at 
around 6~keV.  
However, there have been short term changes in gain due to platform 
temperature changes which are not yet fully calibrated. These changes have 
generally been on timescales of days, due to eclipses, or to changes in 
instrument configuration.
Therefore we analyzed near-contemporaneous PN and MOS-1 data of revolution 59,
(7 days before the observation of {\src}) of the internal in-flight 
calibration source at the location of the cluster using the same responses 
as for the source (MOS-2 data was not usable due to high background). 
We measured the line centroids to be within $\pm^{5}_{15}$ eV 
(or $\pm^{0.1}_{0.2}$ \%) of the 
expected values. This implies a $\pm^{0.0002}_{0.0004}$ systematic 
component of the redshift determination, negligible compared to the 
statistical error of the redshift. Therefore our redshift measurement is 
robust. 

\section{Discussion}

\subsection{The nature of \src}

At a redshift of 0.12, the spatial extent of 2\farcm0 radius 
(see Sect.~\ref{spatext}) corresponds to a physical diameter of 
0.7 h$_{50}^{-1}$~Mpc for the source. 
The only known bound objects on this scale are clusters of galaxies and giant
elliptical galaxies.
The surface brightness profile of {\src} is reasonably well described 
with 2 component $\beta$ model, which is succesfully used to fit the ROSAT 
PSPC data of many cooling flow clusters (Mohr et al. 1999). Our best fit
values for the core radius and the slope are typical of the values in the 
above sample, supporting the cooling flow explanation.

Also, the source has a temperature of 6~keV, typical of rich 
clusters of galaxies. 
The iron abundance is somewhat higher than the typical cluster average, 
but at this redshift the 1\arcmin\ extraction radius 
corresponds to 180~h$_{50}^{-1}$ kpc, less than a typical cluster core 
radius. The recent study of a sample of BeppoSAX clusters (De Grandi \& 
Molendi 2001) shows that in cooling flow clusters the metal abundances 
increase towards the center, reaching values consistent with ours. 
The extraction radius of 180~h$_{50}^{-1}$ kpc is consistent with the cooling 
radius of strong cooling flow clusters (Allen \& Fabian 1997). 
Fitting the spectra of {\src} with an absorbed cooling flow model
{\sc wabs} * {\sc mkcflow} results in a worse fit than the best-fit reported 
above, mainly due to systematic residuals below 2.5 keV. 
This is consistent with the above finding that 30\% of the total emission 
within 1\arcmin\ comes from the ambient gas. However, it's spectral signal 
is too small to be modeled properly. With this scaling, the spectral fit 
indicates a strong mass flow rate of 
$\sim \! 400-600 \ {\rm M_{\odot}}$~yr$^{-1}$, consistent with 
values found for other  clusters (e.g., White et al. 1998). 

For flux calculation, we use the average model normalizations of the PN, 
MOS-1 and MOS-2. 
For the flux uncertainty estimate we consider the full range covered by the 
statistical uncertainties of the model normalizations of all instruments.  
Within a 1\arcmin\ circle centered at the brightness peak, we measure 
unabsorbed 
2--10~keV fluxes and luminosities of
${\rm F_{2-10} = 4.9^{+1.0}_{-0.6} \times \ 
10^{-12}              \ erg~s^{-1}~cm^{-2}}$, and 
${\rm L_{2-10} = 3.5^{+0.8}_{-0.4} \times  \ 
h_{50}^{-2} \ 10^{44} \ erg~s^{-1}}$ 
and bolometric values of
${\rm F_{bol} = 1.0^{+0.2}_{-0.1}  \times   
\ 10^{-11} \ erg~s^{-1}~cm^{-2}}$, 
${\rm L_{bol} = 7.4^{+1.6}_{-0.9} \times 
\ h_{50}^{-2} \ 10^{44}  \ erg~s^{-1}}$.
Using the best fit 2-component $\beta$ model, only 20\% of the total 
emission of the ambient cluster gas is contained within the central 
1\arcmin\ .
For a comparison with L${\rm _{X}}$-T relation determined using ASCA cluster
sample (Markevitch 1998) we
computed the luminosity in 0.1--2.4~keV range within $2 \ h_{50}^{-1}$ Mpc 
using the above $\beta$ model (including the emission from the ambient gas 
and the cooling flow). 
We made no estimate for the uncertainty resulting from the extrapolation of 
the model to larger radius, or from extrapolating the spectral model to the 
absorbed $<$1 keV range. Also, our temperature estimate is not directly 
comparable with the Markevitch (1998) value, because theirs is the 
emission weighted value of the whole cluster, whereas ours is measured in 
the cluster center. Therefore the resulting value 
${\rm L_{X} = 1.5  \times \ h_{100}^{-2} \ 10^{44} \ erg~s^{-1}}$ 
agrees surprisingly well with the corresponding value 
${\rm         1.6 \pm 0.6 \ h_{100}^{-2} \ 10^{44} \ erg~s^{-1}}$ 
derived from the ASCA sample for T = 5.8$\pm 0.6$~keV, with no cooling flow 
correction for T and L$_{X}$. This excellent agreement further supports the 
identification of {\src} with a cluster of galaxies.

The value of \nh\ derived here is 5 times the standard Galactic value of 
$1.6 \ \times \ 10^{22} \ {\rm atom} \ {\rm cm}^{-2}$ obtained from broad 
beam H{\sc i} observations (Dickey \& Lockman 1990).
Fixing \nh\ to the above value gives a totally unacceptable fit to the 
XMM-Newton spectra. The huge excess obtained in our spectral fit, 
$\Delta$ \nh $ \sim 6 \  \ 10^{22} \ {\rm atom} \ {\rm cm}^{-2}$,
is an order of magnitude larger than the typical values in cooling flow 
clusters (e.g., Allen \& Fabian 1997; White et al. 1991). 
This suggests that the excess absorption is due to material in Galaxy, 
rather than in the cluster. The excess could be due to small angular scale 
(\arcmin) variations of \nh\ in the Galactic plane which would be averaged
out in the 1$^{\circ}$ resolution maps of Dickey \& Lockman (1990). To study 
this, we used the dust infrared emission results on a better angular scale 
(6 \arcmin)  (Schlegel et al. 1998),  based on COBE and IRAS all sky maps. 
This data, as computed with Nasa Extragalactic Database extinction calculator,
gives visual extinction A$_{v} \sim 32$ mag in the direction of {\src}.
In a study of ROSAT X-ray halos a correlation between A$_{v}$ and \nh\ was 
found (Predehl et al. 1995), which yields 
\nh $ = 6 \ \times \ 10^{22} \ {\rm atom} \ {\rm cm}^{-2}$ 
for the above amount of extinction.
However, there are several uncertainties involved here.  For example, the 
measurement of A$_{v}$ is not very reliable at low galactic latitudes, and 
the  A$_{v}$ -- \nh\ correlation is measured only in the range 
A$_{v}$ = 0 -- 20 mag. On the other hand a high value of A$_V$ is supported by 
the lack of any obvious cluster glaxy measurements on survey Schmidt plates of the region.
We thus conclude that the \nh\ obtained in this work is consistent with the 
amount of visual extinction, and thus with the actual Galactic \nh.

\subsection{Implications for Galactic plane Surveys}

The ROSAT Position Sensitive Proportional Counter (PSPC) 
8~ks observation of the G21.5-09 field (rp500126n00) shows no
evidence for the presence of \src. This is consistent with the high 
absorption which strongly reduces the flux in the PSPC energy band. 
Consistent with this, a PN image accumulated of the cluster selecting only
photons below 2 keV shows no trace of the cluster.
The 87~ks ASCA observation 50036000 of the G21.5-09 field reveals a weak 
source whose centroid is 3\arcmin\ away from the cluster centroid determined
from the PN image, probably consistent within the uncertainty envelopes. 
However, the source is too faint for any meaningful analysis.
These facts indicate the power of XMM-Newton to enlarge our knowledge
on the galaxy cluster distribution close to the Galactic plane.

De Grandi et al. (1999) based on Rosat All Sky Survey data
from the southern hemisphere provides an estimate of the density of clusters 
as a  function of the ROSAT band flux. In this paper 
we have clearly demosnstrated  that at a  0.5--2.0 keV flux level of 
{\src} within 1\arcmin\ , $\sim 3 \ 10^{-12}$ erg cm$^{-2}$ s$^{-1}$, 
is easily sufficient to measure redshifts accurately 
with XMM-Newton (for reasonable 10--30 ks exposures). 
De Grandi et al. (1999) then predict  60 such  
clusters per steradian. In XMM-Newton A0-1 there are about 10 square
degrees of Galactic plane being covered in systematic surveys, which implies 
$\sim$ 20\% probablility of a detection. 
The above flux is similar to the limit in ROSAT Galactic plane cluster survey 
(Ebeling et al. 2000). Considering that the effective area of XMM-Newton
exceeds that of ROSAT by a factor of 10, and that the exposure times in the 
XMM-Newton surveys exceed those of RASS by a factor of 10, the actual minimum 
flux needed for XMM-Newton cluster detection, with reasonably accurate  
redshift measurement, will be 1-2 orders of magnitude lower than that 
of {\src}. This would 
increase the number of clusters detectable with XMM-Newton on the 
Galactic plane to between a few to tens  per year assuming a similar sky coverage pattern to 
the currently scheduled observations. 

In the Galactic plane galaxy redshifts are very difficult to measure  using
optical techniques due to the large amount of visual extinction and because 
of crowding. We have demonstrated here that XMM-Newton 
has the capability of measuring such redshifts, using the X-ray 
Fe K$_{\alpha}$ line. The systematic uncertainties induce $\sim$ 10 eV uncertainty in the
line centroid measurement which has a negligible effect compared to the statistical
 uncertainties in redshift measurements ($\sim$ 1\%) for clusters with data of similar 
quality and with similar redshifts to {\src}. XMM-Newton data, over its projected 
10 year lifetime will thus prove invaluable, in completing our picture of large scale 
structure in the local Universe and reducing the need for interpolation over a significan fraction of the sky.

\begin{acknowledgements}
Based on observations obtained with XMM-Newton, an ESA science mission
with instruments and contributions directly funded by ESA member states
and the USA (NASA).  J. Nevalainen acknowledges an ESA Research Fellowship.
We thank Dr F. Bocchino, Dr R. Jansen and Dr J. Kaastra for useful discussions. 
\end{acknowledgements}


\begin{thebibliography}{}


\bibitem[]{}
Allen, S., \& Fabian, C., 1997, MNRAS 286. 583

\bibitem[]{}
Allen, S. W., Taylor, G. B., Nulsen, P. E. J., et al., 2001, MNRAS in press, astro-ph/0101162

\bibitem[1989]{a:89}
Anders, E., \& Grevesse, N. 1989, Geochimica et Cosmochimica Acta, 53, 197 

\bibitem[]{}
Aschenbach, B., Briel, U., Haberl, F., et al. 2000, astro-ph/0007256

\bibitem[]{}
Boller, Th., Glozzi, M., Griffiths, M.,  et al. A\&A, 365. L158

\bibitem[]{}
Cavaliere, A., \& Fusco-Femiano, R., 1976, A\&A, 49, 137

\bibitem[]{}
De Grandi, S., B\"ohringer, H., Guzzo, L., et al. 1999, ApJ, 514. 148

\bibitem[]{}
De Grandi, S., \& Molendi, S., 2001, ApJ, in press, astro-ph/0012232

\bibitem[]{}
Dickey, J., \& Lockman, L. F., 1990, Ann. Rev. Ast. Astr. 28, 215. 

\bibitem[]{}
Ebeling, H., Mullis, C., \& Tully, B.,  2000, astro-ph/0001319 

\bibitem[]{}
Fabian, A. C., Sanders, J. S., Ettoriu, G. B., et al. 2000, MNRAS, 318, L65

\bibitem[]{}
Jansen, F., Lumb, D., Altieri, B., et al. 2001, A\&A, 365, L1

\bibitem[]{}
Kraan-Korteweg, R. C., \& Lahav, O., 2000, A\&AAR, 10, 3, 211

\bibitem[]{}
Markevitch, M. 1998, ApJ, 504, 27

\bibitem[1983]{m:83}
Morrison, D., \& McCammon, D. 1983, ApJ, 270, 119

\bibitem[]{}
Predehl, P., \& Schmitt, J. H. M. M., 1995, A\&A, 293, 889

\bibitem[]{}
Saxton, R. D., \& Siddiqui, H. 2000, XMM-Newton technical note. XMM-PS-TN-43

\bibitem[]{}
Schneider, S. E., Skrutskie, M. F., Chester, T. J., et al. 1997, ``Extragalactic Astronomy with 2MASS'', in Extragalactic Astronomy in the Infrared. ed. G. A. Mamon, T. X. Thuan, \& J. T. Thanh Van (Paris: Editions Frontieres), 39

\bibitem[]{}
Schlegel, D. J., Finkbeiner, D. P., \& Davis, M., 1998, ApJ, 500, 525


\bibitem[]{}
Str\"uder, L. Briel, U., Dennerl, K., et al. 2001, A\&A, 365, L18

\bibitem[]{}
Turner, M. J. L., Abbey, A., Arnaud, M., et al., 2001, A\&A, 365, L27.

\bibitem[]{}
Warwick, R. S., Bernard, J.-P., Bocchino, F., et al. 2001, A\&A, 365, L248

\bibitem[]{}
White, D. A., Fabian, A., C., Johnstone, R. M., et al. 1991, MNRAS, 252, 72

\end{thebibliography}
\end{document}